# Structures in the Universe and Origin of Galaxies

V. A. Rantsev-Kartinov

*INF RRC "Kurchatov Institute", Moscow, Russia, rank@nfi.kiae.ru*

**Abstract**

The analysis of images (of obtained in various ranges of the lengths of waves) of luminous objects in the universe by means of a method of multilevel dynamic contrasting led author to the conclusions: **a**) the structures of all observable galaxies represents a complicated constructions which have the tendency to self-similarity and made of separate (basic) blocks, which are a coaxially tubular structures and a cartwheel-like structures; **b**) the majority of observable objects in the universe are luminous butt-ends of almost invisible (of almost completely transparent) of filamentary formations which structures are seen only near to their luminous butt-ends; **c**) the result of analysis of images of cosmic objects show the structure of many pairs of cooperating galaxies point to opportunity of their formation at butt-ends generated in a place of break of the similar filament; **d**) the interacting galaxies (M 81 and M 82) show they are butt-ends of sawed off of two branches of a treelike filament and their interaction is coming out through this filament; e) as our Universe is in dynamics the processes of formation of stars, galaxies and their congestions can go and presently by means of a fracturing of filaments with a corresponding diameters and of the time for their such formation is necessary much less, than along existing standard model.

## 1. Introduction.

Research by author of a skeletal structures of the Universe (**SSU**) began from the analysis of images of various types of plasma by means of a method multilevel dynamic contrasting (**MMDC**), developed and described earlier [**1a, b**]. The analysis of images by **MMDC** is carried out by imposing of various computer maps of contrasting on the image of plasma obtained by the various methods and in any spectral ranges. Some results of the given analysis of a modern database of images of space objects here are given. It is shown, that the topology of the revealed space structures is identical to those which have been already found out and described earlier in a wide range of physical environments , the phenomena and scales [**1-2**]. The basic role in **SSU** is connected with its separate blocks in the form of coaxially-tubular blocks (**CTB**). These **CTB** have complex multi-layered structure of the telescopic enclosed tubes which lateral walls represent a weaving of similar filaments of smaller diameter, with the central cord. Except for that, these blocks inside have also radial connections. Extended filaments of these structures are collected of almost identical **CTB**, which are flexibly connected among themselves as in joints of a skeleton. It is assumed such joints may be realized due to stringing of the individual **CTB** on common flow of the magnetic field which penetrates the whole such filament, and itself the **CTB** are the interacting magnetic dipoles with micro-dust skeletons, which are immersed into plasma.

## 2. Observations of cartwheel structures in laboratory and cosmic plasma

Here, we will try to connect laboratory experiments and space by considering a short gallery of cartwheel-like structures, which are probably the most inconvenient objects for universally describing the entire range of observed space scales. In laboratory electric discharges [1] and their respective dust deposits [2] the cartwheels may be located either on the butt-ends of a tubes, or on an «axle-tree» filament, but they may be separate blocks also (the smallest cartwheels are of diameter less than 100 nm (see Figs. 2 and 3 in *Phys. Lett. A,* **26**9, 363-367 (2000)). So, similar structures of different scales are found in the following typical examples: (i) big icy particles of a hail (Fig. 1A), (ii) a fragment of tornado (Fig. 1B), (iii) a supernova remnant.

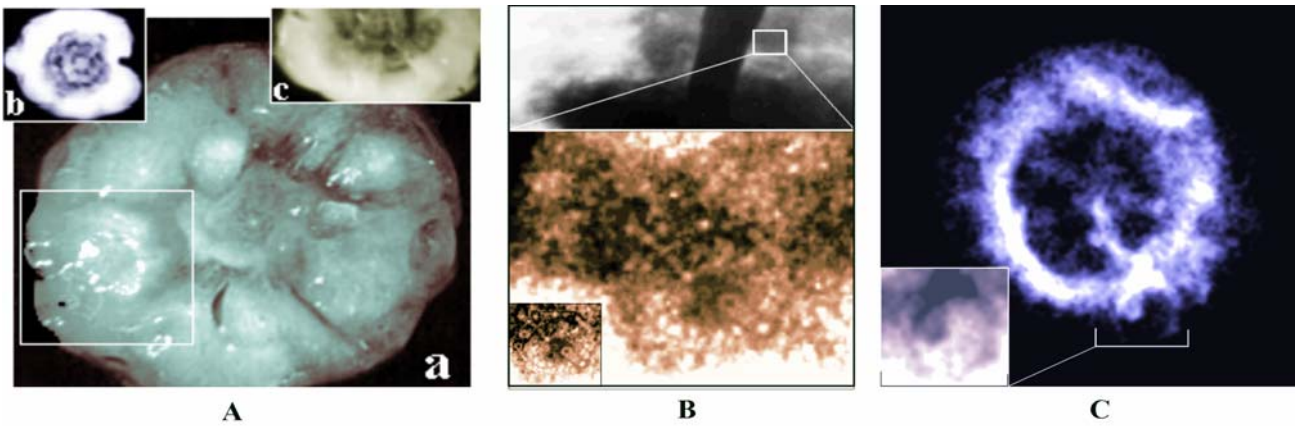

**Fig. 1.** The cartwheel-like structures at different length scales are presented. **A)** Big icy particles of hail of diameter 4.5 cm (a), 5 cm (b), and 5 cm (c). The frame in the left lower part of the image (**a**) is contrasted separately to show an elliptic image of the edge of the radial directed tubular structure. The entire structure seems to contain a number of similar radial blocks. A distinct coaxial structure of the cartwheel type is seen in the central part of image (**b**). Image (**c**) shows strong radial bonds between the central point and the «wheel». **B)** Top section: A fragment of the photographic image of a massive tornado of estimated size of some hundred meters in diameter. Bottom section: A fragment of the top image shows the cartwheel whose slightly elliptic image is seen in the center. The cartwheel seems to be located on a long axle-tree directed down to the right and ended with a bright spot on the axle's edge (see its additionally contrasted image in the left corner insert on the bottom image). **C)** «A flaming cosmic wheel» of the supernova remnant E0102-72, with «puzzling spoke-like structures in its interior», which is stretched across forty light-years in Small Magellan Cloud, 190,000 light-years from Earth [13]. The radial directed spokes are ended with tubular structures seen on the outer edge of the cartwheel. The inverted (and additionally contrasted) image of the edge of such a tubule (marked with the square bracket) is given in the left corner insert (note that the tubule's edge itself seems to possess a tubular block, of smaller diameter, seen on the bottom of the insert).

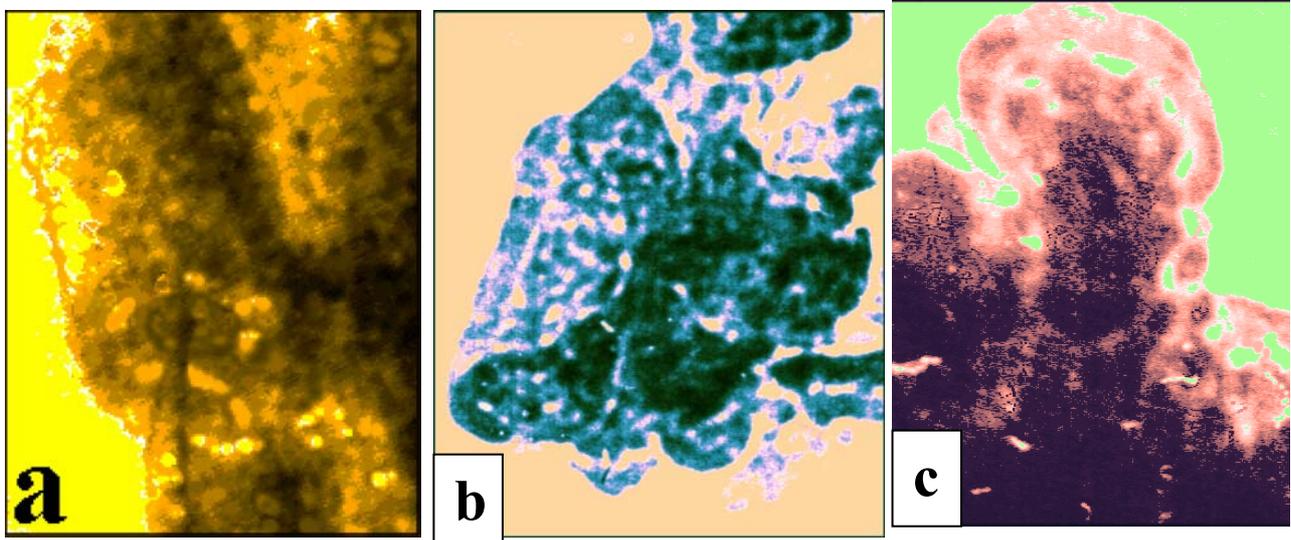

**Fig. 2. a)** The «cartwheel» structure in tokamak TM-2 plasma (minor radius 8 cm, toroidal direction is horizontal)  Diameters of larger and smaller ring-shaped structures on a common axle are ~2.2 cm and ~1 cm, respectively.  Image (positive) is taken in visible light by a streak camera with time resolution < 1 µs (original picture is taken from Vinogradova N.D., Razumova K.A. Int. Conf. Plasma Phys., Culham, U.K., 1965). **b)** The TEM image (magnification 34,000) of a small fragment of dust particle (1.2 µm diameter) extracted from the oil used in the vacuum pumping system of tokamak T-10. The tubule, whose edge with the distinct central rod is seen in the lower left part of the figure, is of ~70 nm diameter and ~140 nm long. Diameter of slightly inhomogeneous cylinder on the left side of the tubule is ~10 nm. The radial bonds between the side-on cylinder and the central rod are of ~ 10 nm diameter. **c)** Another fragment of the same dust particle. The cartwheel (a coaxial two-ring structure, D ~ 100 nm) is declined with respect to image's plane and located on a thick rod (which probably "works" as an axle of the cartwheel).

Note that the cosmic wheel's skeleton (FIGURE 1C) tends to repeat the structure of the icy cartwheel (FIGURE 1A) up to details of its constituent blocks. In particular, some of radial directed spokes are ended with a tubular structure seen on the outer edge of the cartwheel. Moreover, in the edge cross-section of this tubular structure, the global cartwheel of the icy particle contains a smaller cartwheel whose axle is directed radially (see left lower window in FIGURE 1A). Thus, there is a trend toward self-similarity (the evidences for such a trend in tubular skeletons found in the dust deposits).

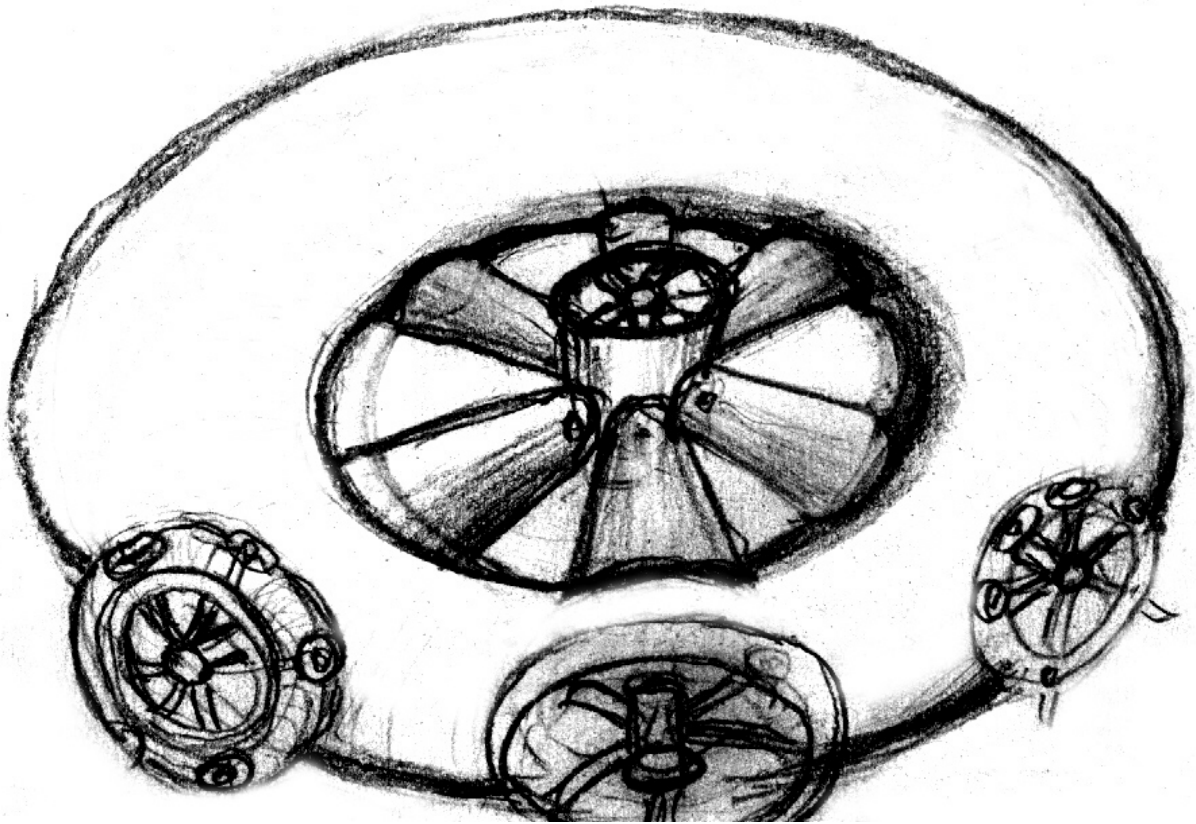

**Fig. 3.** The schematic image of structures such as "cartwheel" is given here. Thus, the cosmic wheel's skeleton tends to repeat the structure of the cartwheel itself up to details of its constituent blocks as in the icy cartwheel (see Fig. 1A).

### 3. Observations of "electric torch-like structures" in laboratory and cosmic plasma

One of the new phenomena which has been found out at the analysis of images of plasma, were a rectilinear dark filamentary structures which butt-end can shine as open butt-end of optical fibers in such ranges of lengths of waves which correspond with temperature of researched plasma. Such the **CTS** have been described and have received the name" electric torch-like structures" (**ETSs**) [**1c, d**].

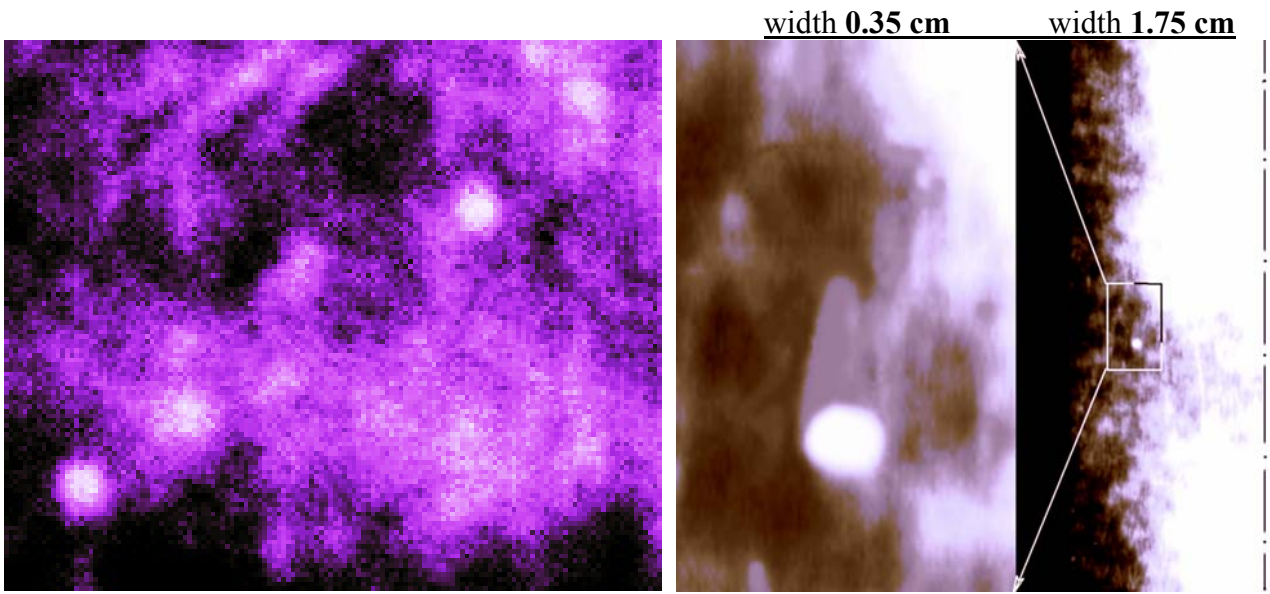

**Fig. 4. a**) The image of a luminescence of tokamak T-6 of plasma in seen light, at exposure ~ **1 μ s**. The height of a picture corresponds to ~ **8 cm**. In the left bottom corner the tubular structure such as **ETS** with diameter ~**1 cm** is visible and directed almost along a diagonal of figure. **b**) Right section: The image (width **1.75 cm**) of the left-hand side of the denser and hotter core of the vertically aligned long plasma column (the Z-pinch axis in shown with a dashed-and-dotted line) in the electric discharge of Z-pinch type. The image (positive) is taken in the visible light with time exposure **2 ns** nearly at the moment when discharge's magnetic field squeezes the hot plasma column and partly strips a skeletal network from ambient luminous plasma (for an example of a strongly stripped skeleton in the same Z-pinch. Left section: The magnified, **0.35 cm** wide window reveals the «hot spot» to be the edge of the filament which is close enough to a brighter core.

        **A**                        **B**                        **C**

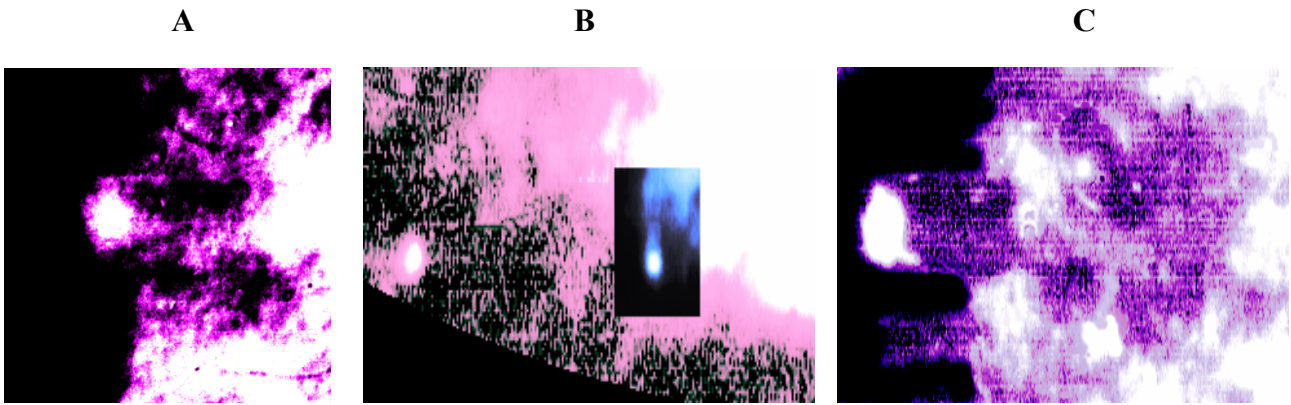

**Fig. 5.** The"electric torch-like structures" (ETLS) in Z-pinch plasmas, electronic temperatures ~ 300 eV, density ~ $10^{18}$ cm$^{-3}$, time of exposure ~ 2 ns. Heights of these images are ~ 1 mm, diameters of observable ETLSs are ~ 0.25 mm.

    It seems that the straight blocks of skeletons may work as a guiding system for (and/or) a conductor of electromagnetic signals. Therefore, the open end of a dendrite electric circuit or a local disruption of such a circuit (e.g., its sparkling, fractures, etc.) may self-illuminate its and to make its observable. The similar phenomena are observed in space plasma*.* ***Many luminous objects in the Universe represent such luminous butt-end of the CTS.*** The butt-ends of such open optical paths can correspond to sizes of stars, planetary nebulas, or galaxies and their congestions.

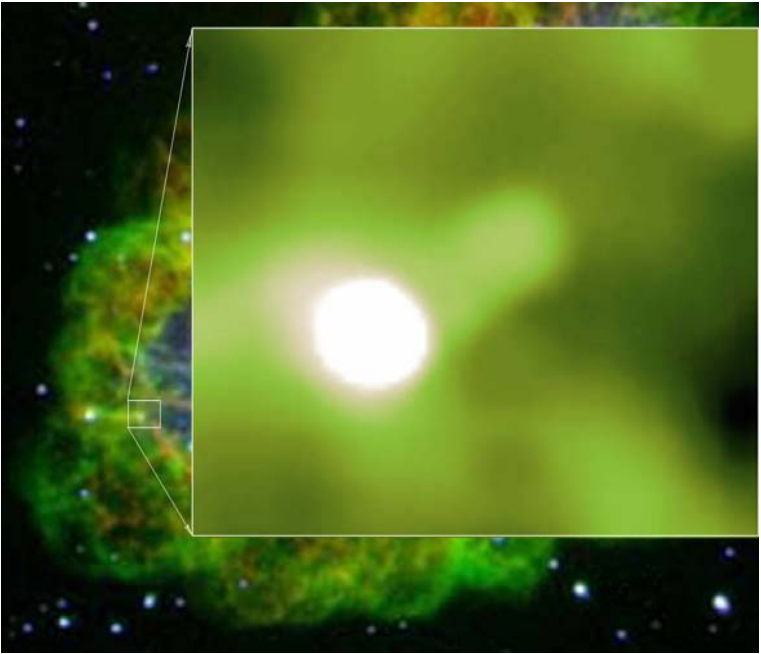

**Fig. 6.** The torch-like structure of the filament of ~ 0.1 light year in diameter is seen in the fragment of the optical image of the Crab nebula [**3**]). The light from the filament's edge anisotropically illuminates the ambient gas. Regardless the cylindrical straight body of the filament is visible due to internal radiation source or external illumination, the visible continuity of the filament is not destroyed by the partly opaque surrounding medium (seen in the left upper corner of the window) and may be traced in the full image outside the window.

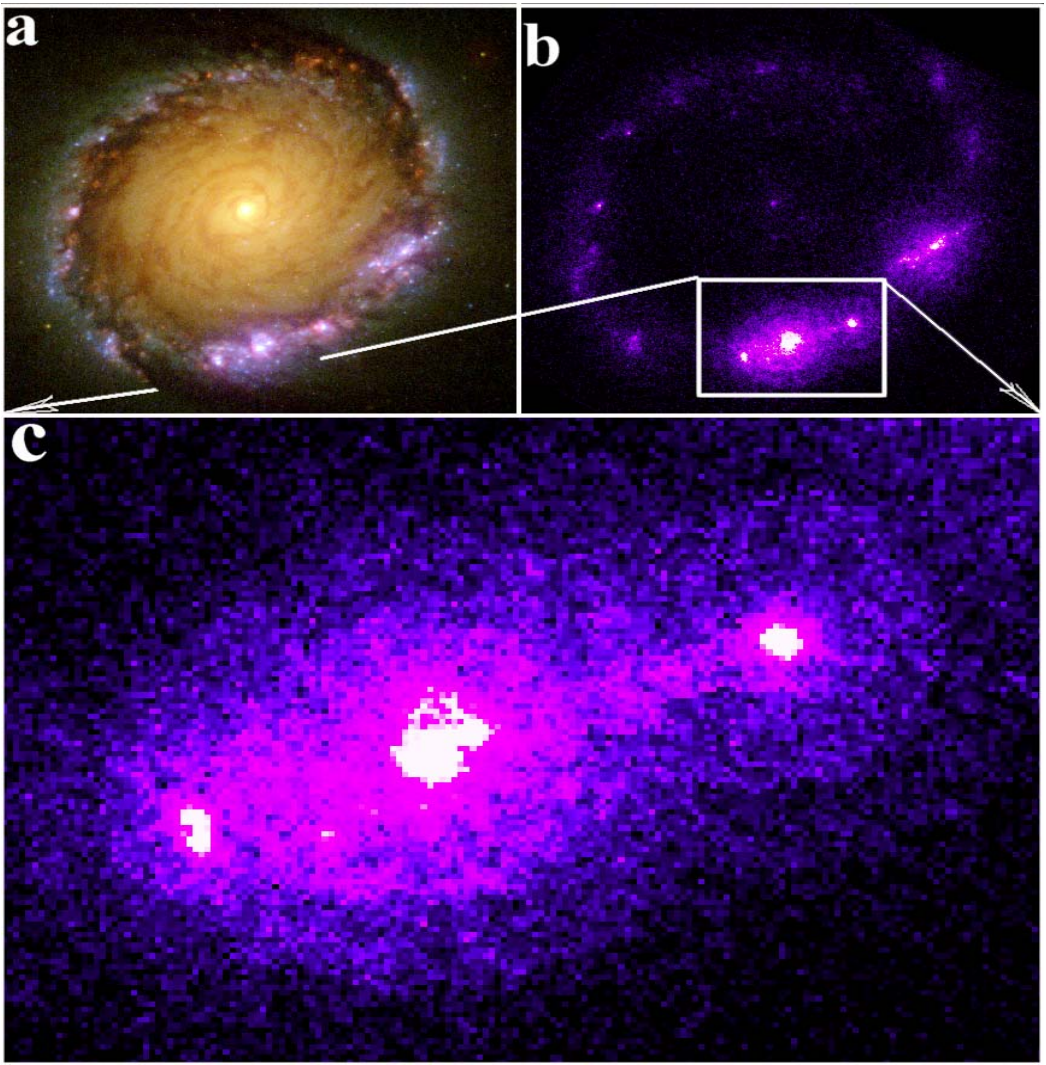

**Fig. 7.** The electric torch-like structure in the **2,400 light-year-wide core** of the spiral galaxy **NGC 1512** [**4**], which is **30 million light-years** away and **70,000 light years across**. (**a**) Color-composite image of the core, (**b**) similar image in the **2200 Å light**, (**c**) contrasted image of a part of the image «b» (diameter of straight filament with a bright spot on its edge, seen in the upper right, is ~ **60 light-years**).

# 4. Revelation of a Coaxially-Tubular Blocks of the SSU.

The **CTS** are basis of the **SSU** as this type of blocks composes an overwhelming part of it. It is possible to show, that almost all luminous objects in the Universe are butt-ends of the above mentioned blocks of either scale. The **CTS** revealed in the Universe tend to self-similarity that leads to fractality of built by them and by observed structures. It is easy to demonstrate it by the examples of a structure of blocks of various scales of galaxies. The central fragment very large spiral galaxy where are precisely revealed the **CTS** of various scales (from $10^{17}$ **cm** up to $4 \cdot 10^{18}$ **cm**) as an example is produced in the next figure.

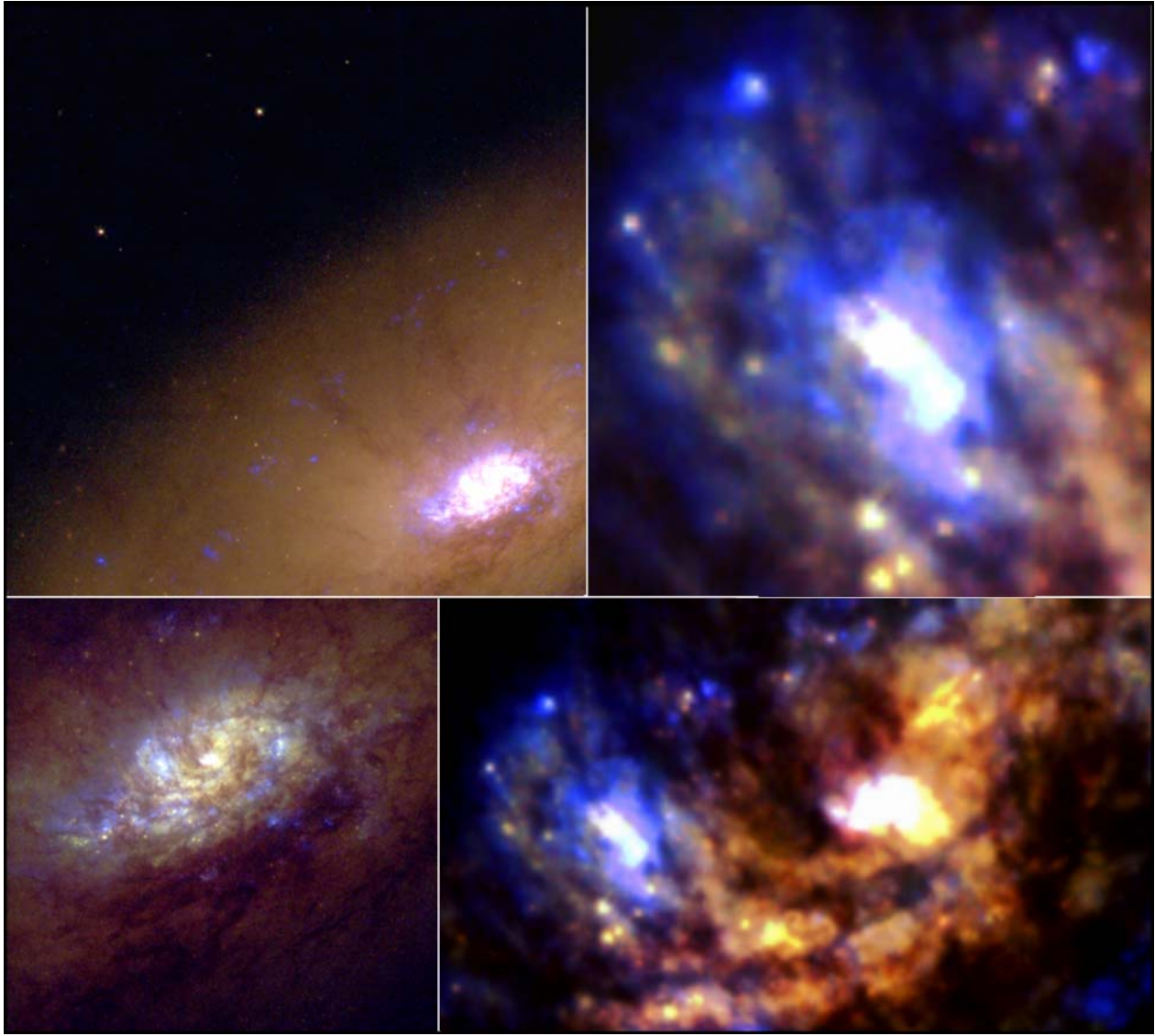

**Fig. 8.** **a)** Spiral galaxy NGC 1808. Constellation: Columba **[5]**. Distance: **40 million light years.** **b)** A central portion of this galaxy. **c)** The width of figure corresponds ~ $7 \cdot 10^{18}$ **cm**. The arrow specifies a direction of an axis of a galaxy. In the center of this galaxy it is revealed: the **CTS** (cylinder) in diameter ~ $2.3 \cdot 10^{18}$ **cm** ; a telescopic putted tubular block (by diameter ~$1.7 \cdot 10^{18}$ **cm** ) on its axis; a dark filament (in diameter ~ $5 \cdot 10^{17}$ **cm**) inside of them on axis of galaxy. Continuation in space above a galaxy of a central filament is a dark filament, leaving on the center same filament, but the greater diameter. The similar structure from telescopic enclosed the **CTS** is revealed to the left of the axis. It has of blue the **CTS** in the center of the similar structure built from tubes with a bright blue luminescence of their butt-end. In the center inside this structure it is precisely looked through bright white the **CTS** which center leaves dark thin filament which butt-end above all structure shine a blue luminescence. **d)** The magnified part of the window "**c**".

## 5. Revelation of skeletal structures of the cooperating galaxies

The **CTS** are most representative in the common structure of the Universe. Their participation in its dynamics leads to observation and definition of role of these structures in every possible space accident. It is possible to show, *all cooperating galaxies represent interactions of the similar CTSs in the form of a breaks and/or collisions of the CTSs of a corresponding sizes.* The image processed by means of the **MMDC** and received in the field of ultraviolet on next Fig. 5 is given, which represents the image of two cooperating galaxies, **UGC 06471** and **UGC 06472**, and corresponding schematic of its representations in a window on the right.

http://hubblesite.org/newscenter/newsdesk/archive/releases/2001/04/image/d

**The galaxy UGC 06471 and UGC 06472**

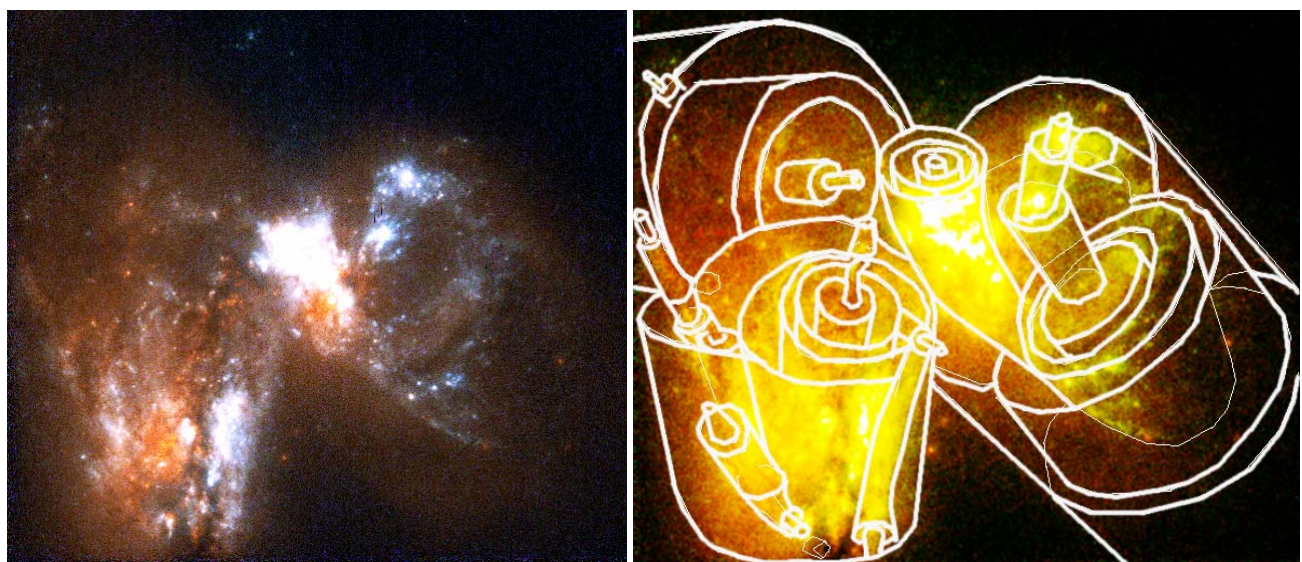

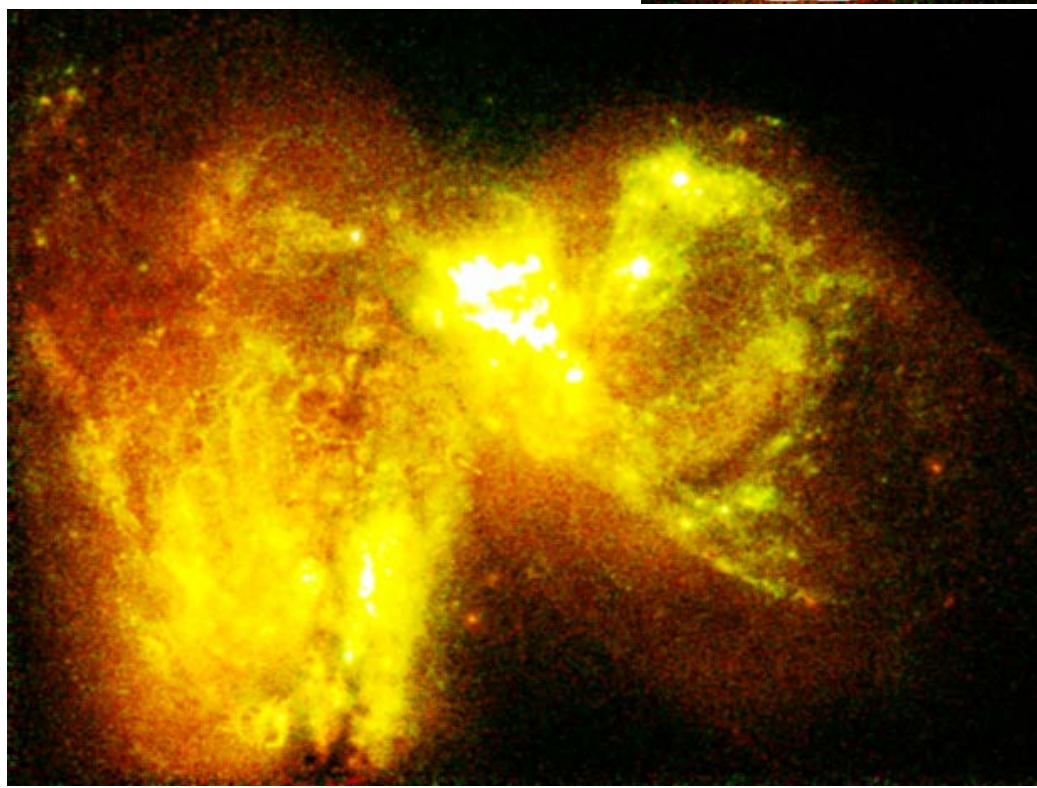

**Fig. 9.** The galaxy UGC 06471 and UGC 06472 [6]. It is revealed, that the given object represents interaction of a three CTS. It is possible to assume, that two galaxies in the foreground are butt-end of a break of one the CTS (from the analysis of identity having chopped off on their external environments).

The third CTS could be the reason of such break, as a result of its collision with a joint of first two. More thin analysis of this group of the CTS shows their topological identity. These blocks represent telescopic the enclosed tubes with radial connections.

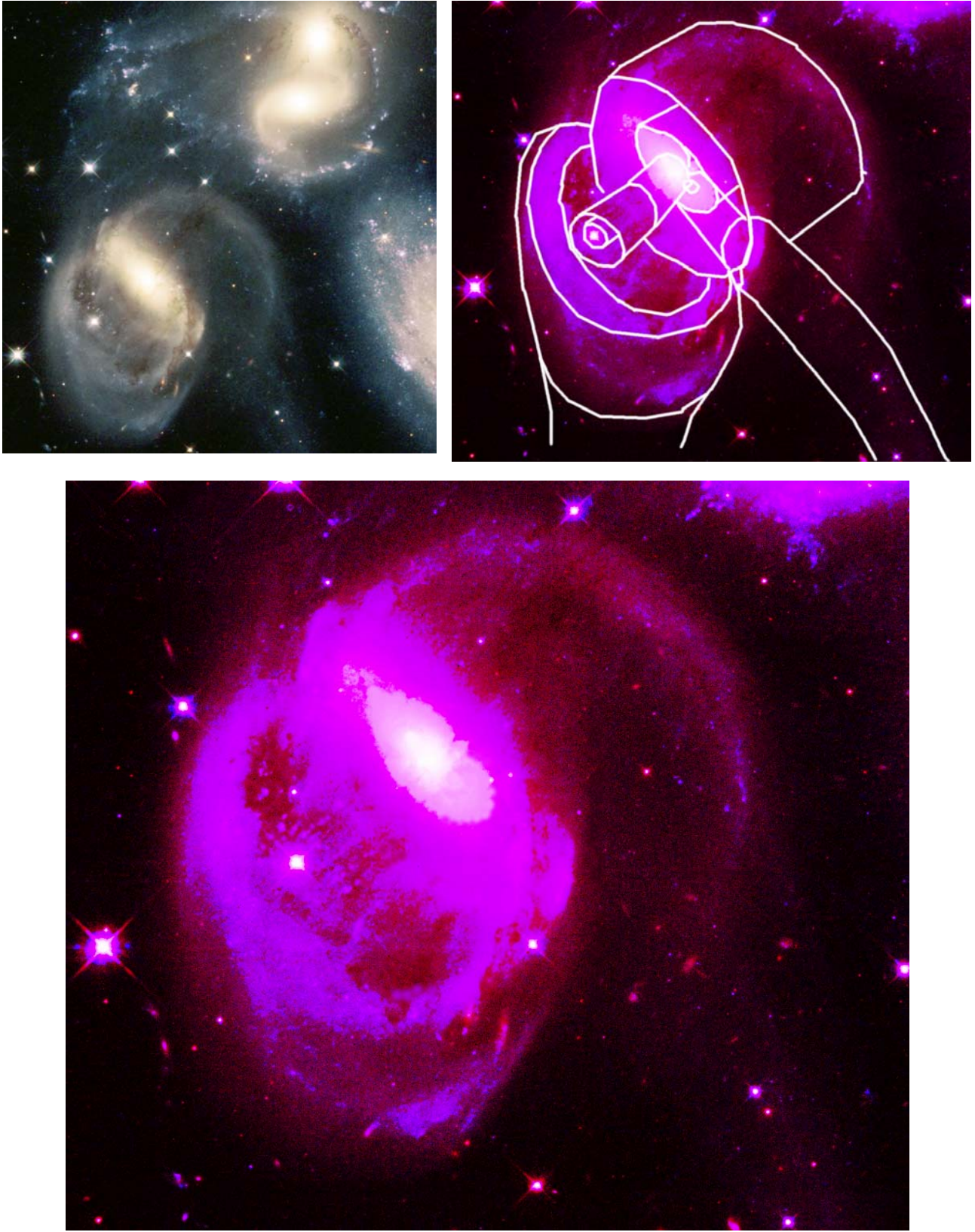

**Fig. 10.** Two interacting galaxies **NGC 7318 A** and **NGC 7318 B** [7] which are in constellation Pegasus on distance **270 million light years** from the Earth are given here. These galaxies represent mutually - perpendicular butt-ends of a break of one **CTS** with diameter ~ **3 $10^{22}$ cm.** The structure of this break is precisely traced and its plan is given in a window below. This structure has fractured in a place of its fold at folding up twice. The multi-layered design of an internal structure of break-up of this **CTS** is very well seen. These galaxies were born as a result of a given filament fracturing.

A

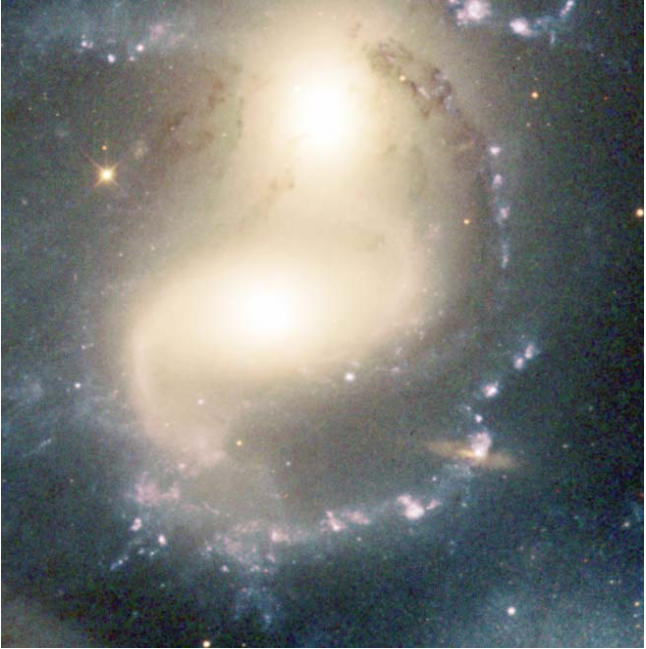

B

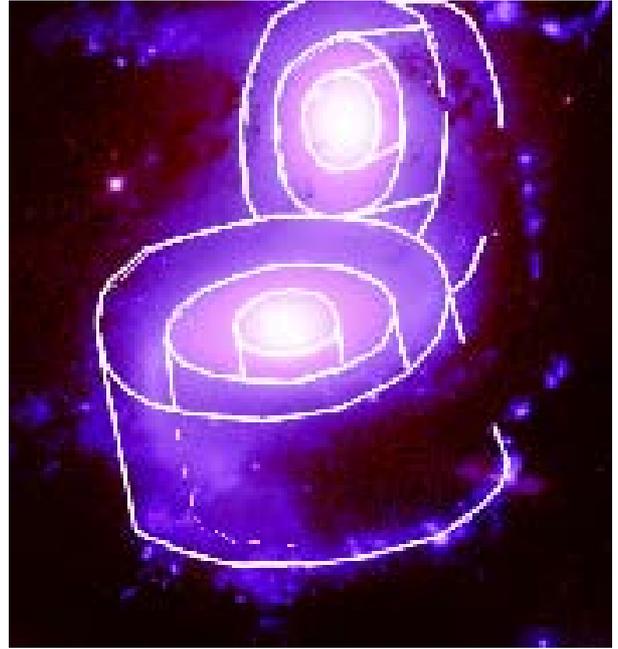

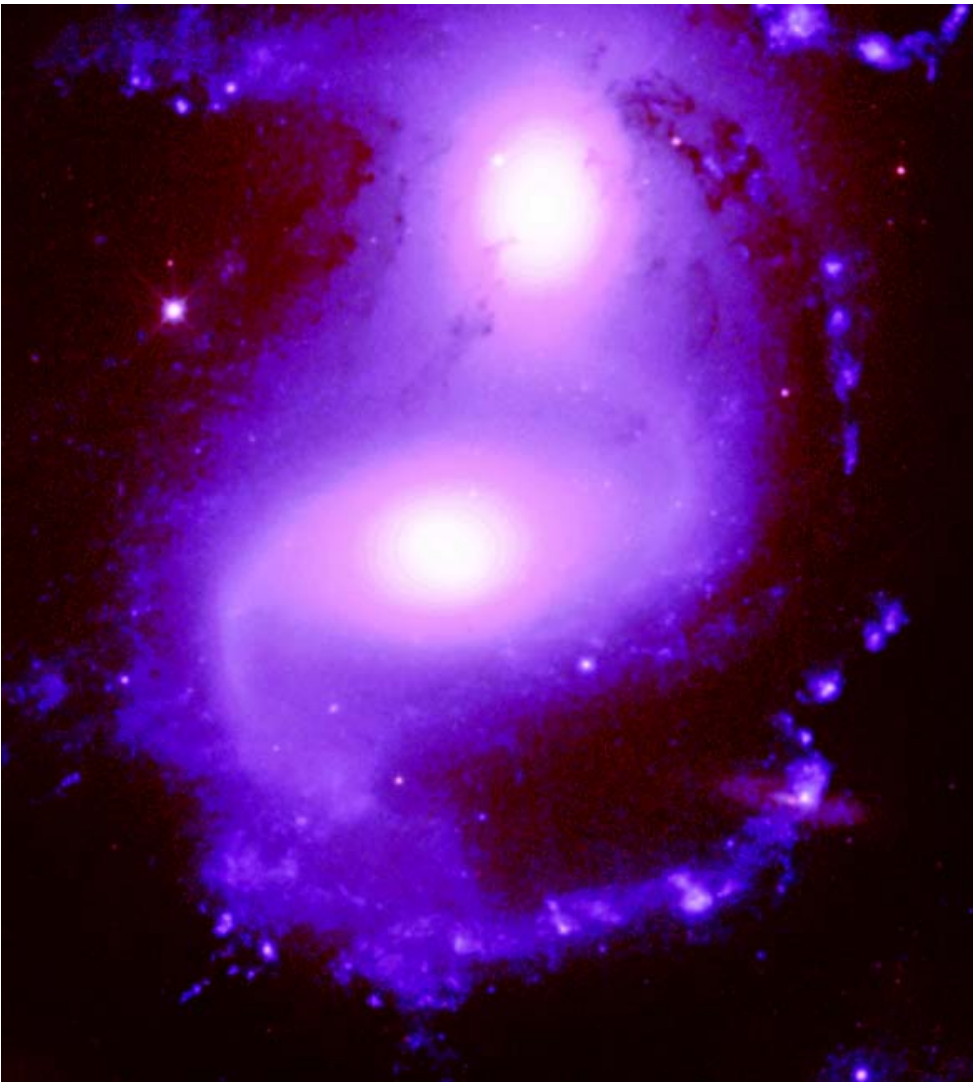

**Fig. 11.** **A)** Here we have the cooperating galaxies **NGC 7319A** and **NGC 7319B** [7] of the same constellation Pegasus which also are by result of a fracturing of the **CTS** with diameter ~ **3 $10^{22}$ cm**. **B)** Here the result of the **MMDC** analysis of this image is presented. Butt-ends of the fracturing are located mutually perpendicularly. Dark filament on the right shows possible points of interface of the fracturing. From under index **"b",** at the left, the outline of a dark filament which could be the reason of fracturing of this **CTS** is seen. **C)** This is a scheme of the image.

C

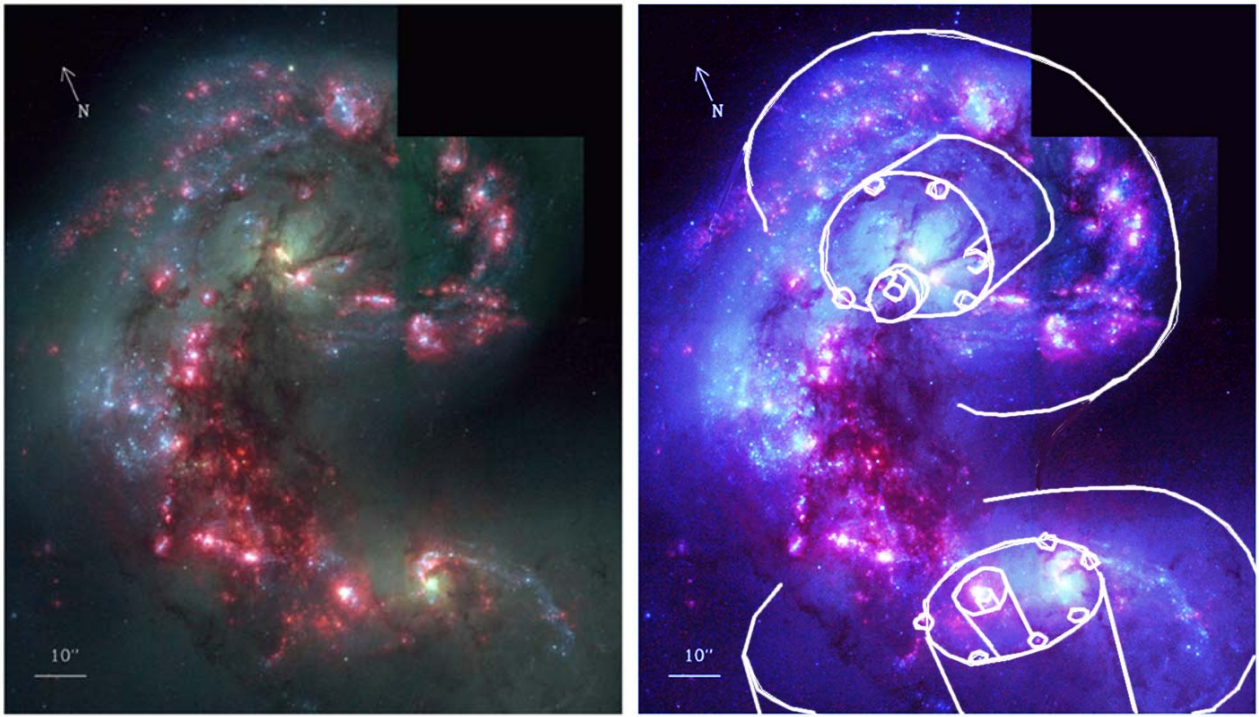

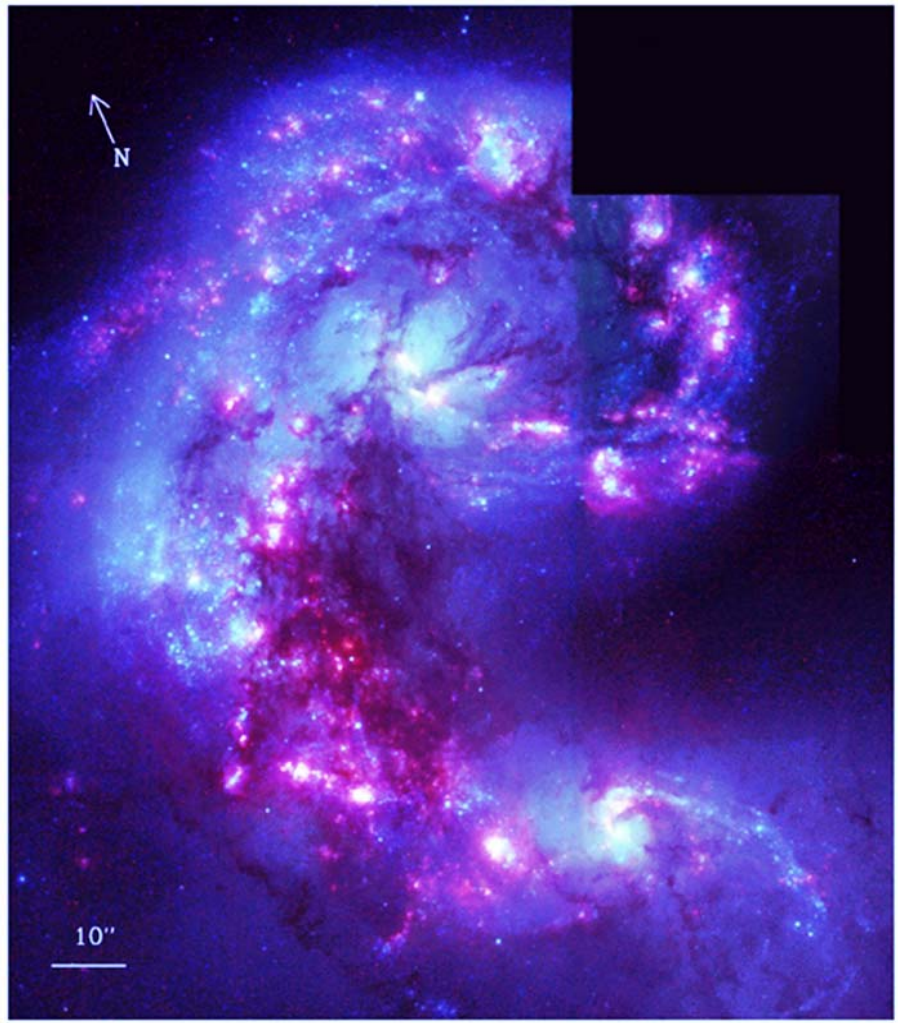

**Fig. 12. a**) The images of two cooperating galaxies NGC 4038/4039 of constellation Corvus [8] are submitted. **b**) Here the result of the MMDC analysis of galaxies image is presented. Identity of structures of these two galaxies as topological, so and in the sizes is visible. Moreover, in structure of the top galaxy a radial spokes which connects of a central coaxially-tubular part of galaxy with tubes located on lateral surface of a filament with the greater size are looked through. It is seen, what interface details of the lower galaxy have the same sizes as at the upper galaxy external coaxial of the top galaxy shows that the coaxially-tubular structures from which it consist have the same topology as the topology of whole galaxy. These two galaxies as a punch and a matrix have the same topology and the size. That is, their birth is caused by a breach of one CTS because their corresponding elements can be combined precisely. **c**) This is schematic representation of the given image.

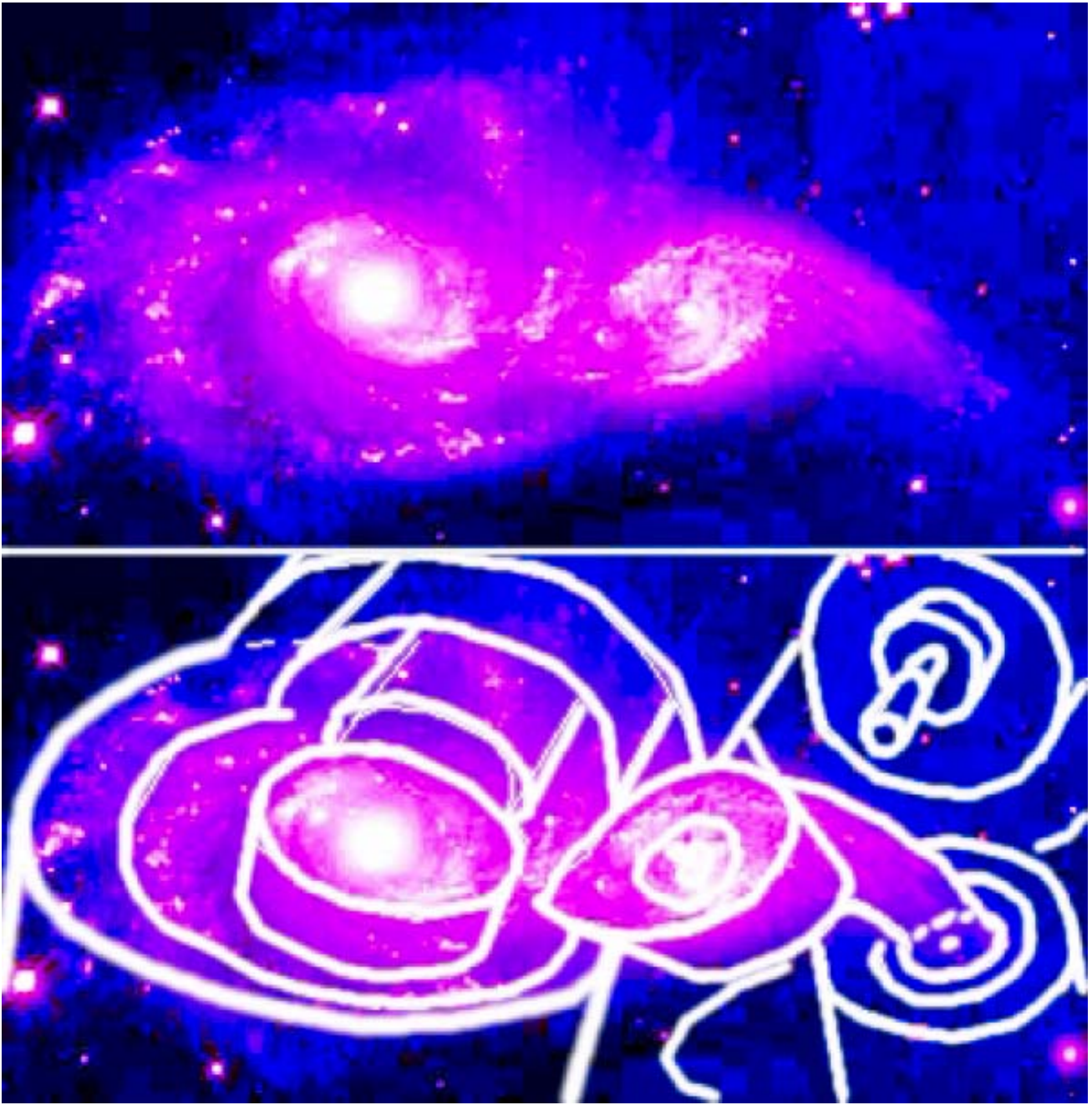

**Fig. 13.** Here two cooperating galaxies - NGC 2207 and **IC 2163** are submitted [**9**]. The width of figure corresponds ~ **4 10$^{23}$ cm**. The analysis of the image (by means of **MMDC**) is applied for revealing structure of interaction. For simplification of perception the schematic image of the given interaction of galaxies below is given. It is visible, that the given galaxies represent butt-ends of the broken **CTS** (the larger galaxy, **NGC 2207**, is on the left; the smaller one, **IC 2163**, is on the right). The first of them has the same size which is characteristic for spiral galaxies ~ **2 10$^{23}$ cm** Proceeding from the submitted scheme, it is possible to tell that the script of observable process can appear absolutely other, than it is represented now by astronomers.

The large-scale CTS, incorporating among themselves can build a united network of the Universe. In that case all objects of this network are in direct connection. Such communication is shown for the neighboring galaxies especially strongly when they belong to one treelike filament, being butt-ends of its branches. Sometimes this is able to reveal, and demonstrate.

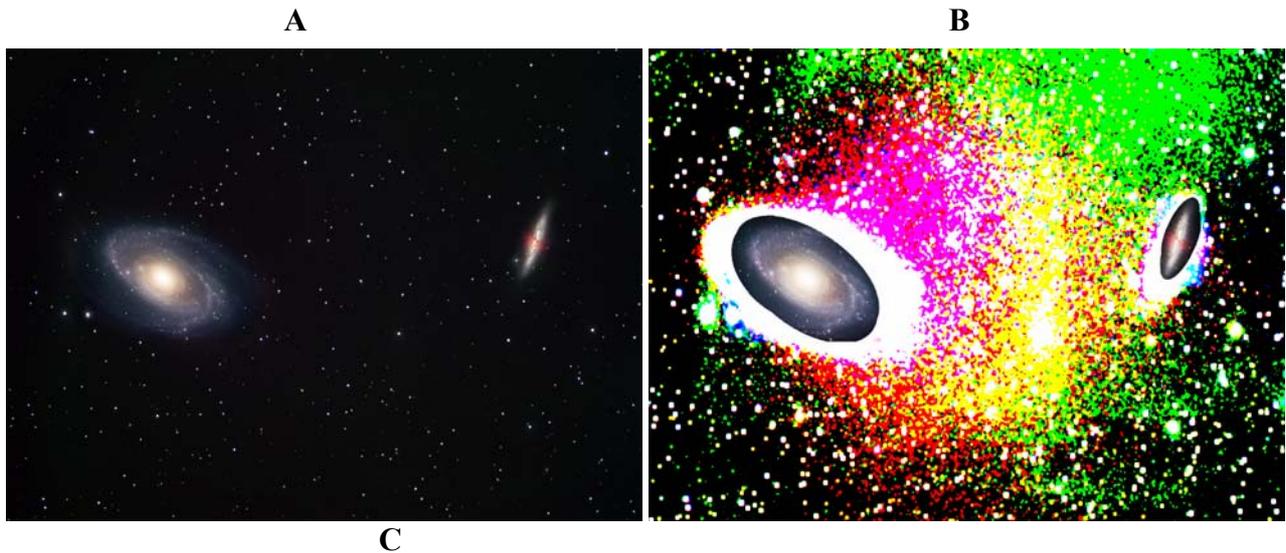

**Fig. 14.** **A**) They are the cooperating galaxies M81 and M82 of the constellation « the Big she-bear» [10]. **B**) The MMDC has allowed revealing their structural interaction. M81 is a butt-end of an internal and acting part of a treelike filament with diameter ~ 3.5 $10^{23}$ cm, and M82 a butt-end of its lateral cut down branch. To the right of the M82 it is looked through parallel to an axis of the basic filament the dark CTS which, obviously, has cut off a branch, having created the given galaxy. The bright luminous object located before M82 and hardly below of it, lays on an axis of this dark structure. **C**) This is a schematic representation of the image. ***Contrary to opinion of contemporary astronomers, according to the image, obviously, these galaxies did not collide.***

## 6. Cartwheel-Like Structures of the Universe

The CWS are the most interesting and complex observable blocks in the Universe, and also they are the most typical blocks of the observable fractal structures of which are difficultly to confuse with any another. If such structure is well oriented in a flatness of a shearing, then (at condition of a corresponding statistics) the structure clearly becomes apparent because the basic massif of points of this structure is fitted to a rim of a wheel, to its axis and radial spokes, making (on the square) half of area of whole wheel. It allows to identify precisely its under such circumstances.

It is theoretically difficult to explain topology CWS by means of magnetic hydrodynamics. The mechanism [1a,b] of construction of the revealed by us topology of fractals spontaneously gathering at formation of electric breakdown at the presence of elementary blocks of a dust, which have tendency to forming structures (for example, as carbon nanotubes or a similar structures but of other elements and chemical compounds) have been earlier considered. The sequence of generations CWS right up to the size ~ $10^{23}$ cm already has been shown earlier [1a, b]. The result of the analysis by means of MMDC of maps of redshifts is given below.

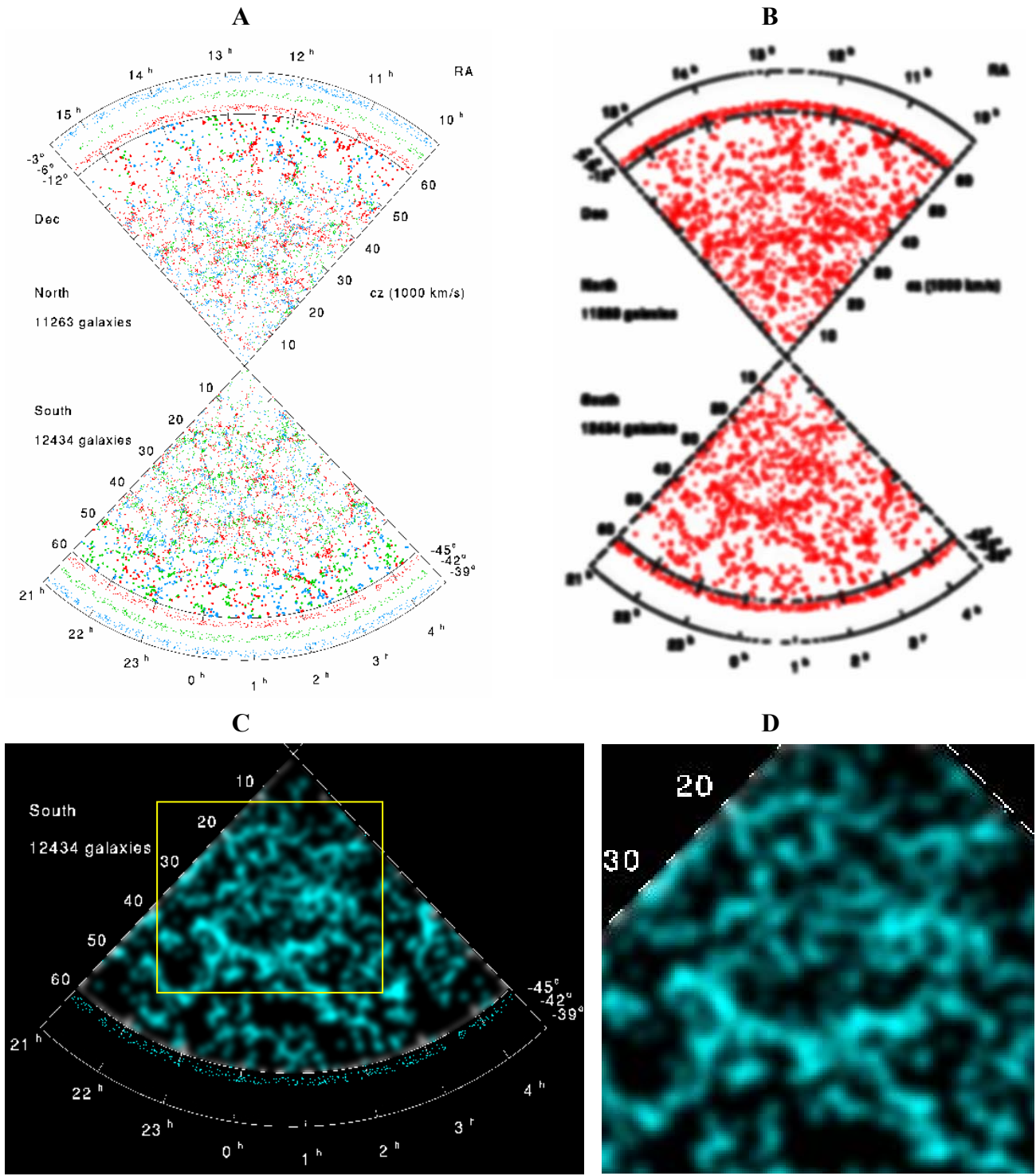

**Fig. 15. A)** A fragment of distribution of the galaxies (20,000 galaxies (for redshifts Z < 0.3, i.e. at distances L up to 2.5 billion light-years away) 3.0° thickness of slice is centered at declination - 45° in the South Galactic Pole strip. **B)** The image of map of redshift in the red points in the colored image at http://www.astro.ucla.edu/~wright/lcrs.html [11]. The left border of the cone crosses the left hand side of the figure at a distance ~ 1.5 $10^9$ light years. Thickening of the spots with subsequent smoothing of the image are used here for correlation analysis the result of which gives a circle and a straight radial filaments - connections similar to a spokes. **C)** Here the increased image of South Pole of a redshift map after carrying out of the correlation analysis is given. At the left it is clearly seen the structure of type of cartwheel. **D)** Here the increased image of window in image C is given. It is clearly seen the components of cartwheel structure, the diameter of which is about 1.5 $10^{27}$ cm (1.5 $10^9$ of light years).

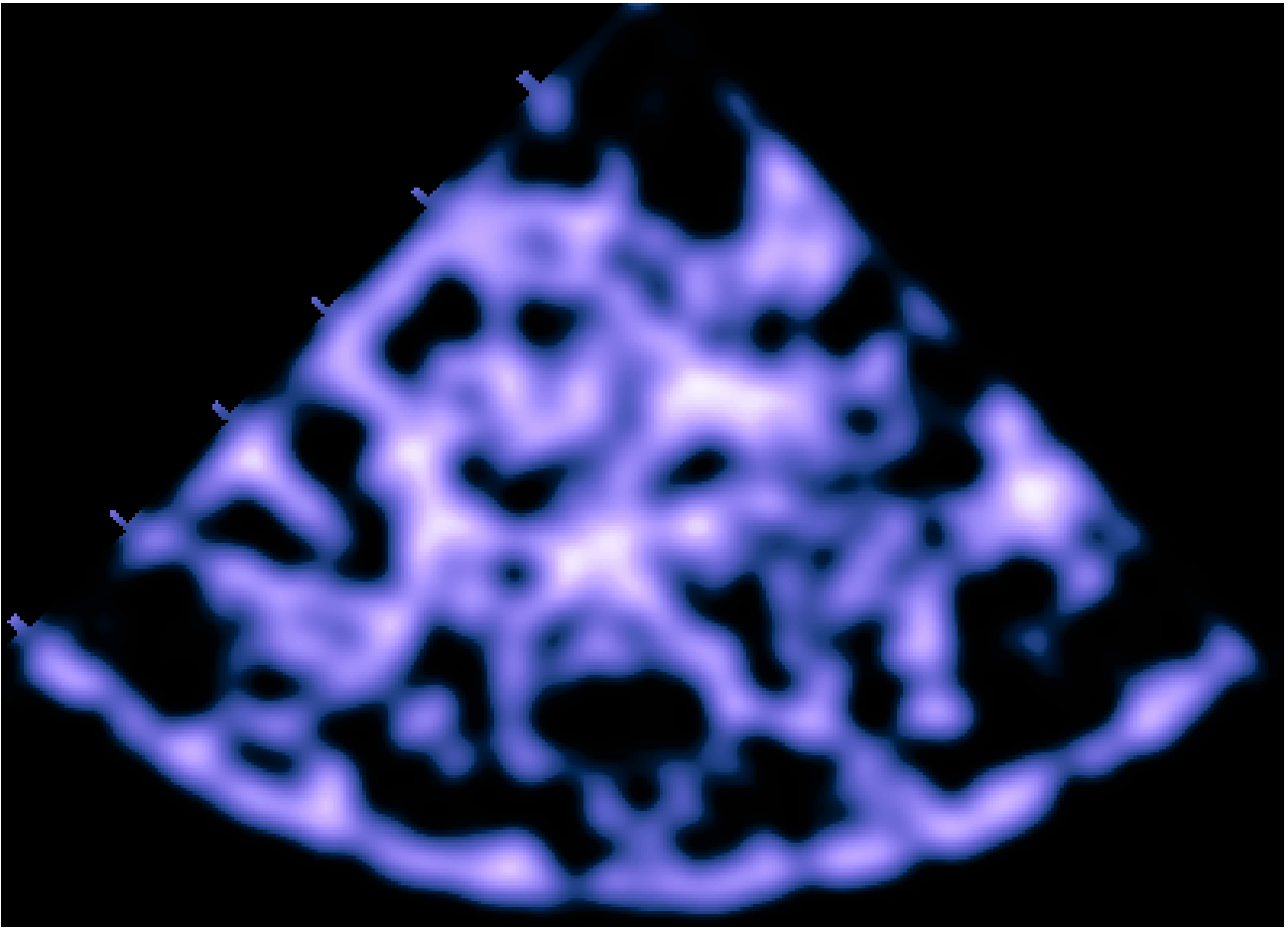

**Fig. 16.** Here have given the revealed Universe structure on the Southern part mentioned above databases, but not on 3.0 degree a cut but on its full set, i.e., its 9 degree a variant.

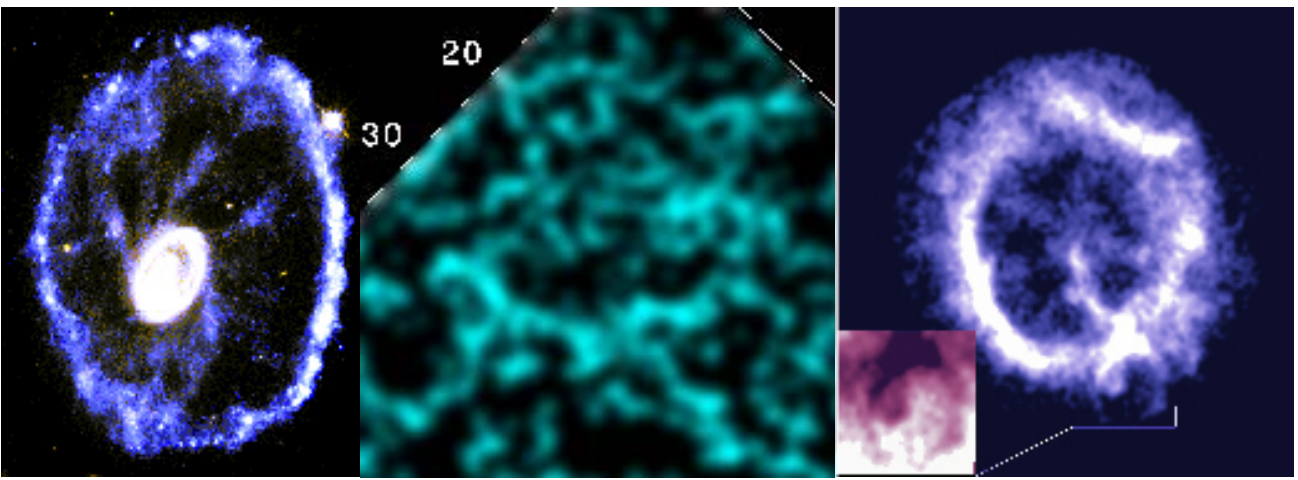

**Fig. 17.** In center it is given image of the CWS with scale ~ 1.5 $10^{27}$ cm which is result of analysis by means of the MMDC of a map redshift [11], at the left it is given image of the CWS of galactic scale (~ $10^{23}$ cm) [12]; on the right it is given image of the CWS of an explosion of supernova (~ 4 $10^{19}$ cm) [13]. Here topological identity of observable large-scale structures of the Universe distinguished almost on 8 orders of size is precisely traced. It is well visible, that on an exit of radial spokes through a rim of the CWS it is formed structures similar to it itself. Through them the given structure is intertwined into the general network of the Universe.

# 7. CONCLUSIONS

a) *Our Universe has general skeletal structure which is made of separate coaxially-tubular blocks and blocks of type cartwheels.*
b) It is self-similar on any scales, and, hence, fractal.
c) <u>All luminous objects observable in the Universe are or by free butt-ends of the corresponding sizes of coaxially-tubular blocks (of telescopic putted in each other), or by breaks of filaments which is assembled of these blocks.</u>
d) As our Universe is in dynamics ***<u>the processes of formation of stars, galaxies and their congestions can go and presently by means of a fracturing of filaments with a corresponding diameters and of the time for their such formation is necessary much less, than along existing standard model.</u>***

      This is evidently shown on an examples of revealed structures at the analyses of the images of cooperating galaxies and a part of structure of the Universe showing the same topology which has been revealed by us on a wide range of scales and the phenomena earlier. The identity of observable topology now is shown from $10^{-7}$ cm up to $10^{28}$ cm, i.e., on 35 orders of size. It is possible to assume, that *if we will be able to analysis of structure of substance at moving aside reduction of scales there we shall discover the same topology*. That it means and what will be a result for us - will show time.